\documentstyle[prd,aps]{revtex}
\begin{document}
\input epsf
\draft
\renewcommand{\topfraction}{0.8}
\twocolumn[\hsize\textwidth\columnwidth\hsize\csname
@twocolumnfalse\endcsname \preprint{CITA-99-12, SU-ITP-99-13,
hep-ph/9903350,   March 14, 1999}
\title { \Large  \bf  Inflation and Preheating in NO models}
\author{Gary Felder,$^1$ Lev Kofman,$^2$ and Andrei Linde$^1$}
\address{${}^1$Department of Physics, Stanford University, Stanford, CA
94305, USA}
\address{${}^2$CITA, University of Toronto, 60 St George Str,
Toronto, ON M5S 1A1, Canada}
 \date{March 14, 1999}
\maketitle
\begin{abstract}
We study inflationary models in which the effective potential of
the inflaton field does not have a minimum, but rather gradually
decreases at large $\phi$. In such models the inflaton field does
not oscillate after inflation, and its effective mass becomes
vanishingly small, so the standard theory of reheating based on
the decay of the oscillating inflaton field does not apply. For a
long time the only mechanism of reheating in such non-oscillatory
(NO) models was based on gravitational particle production in an
expanding universe. This mechanism is very inefficient. We will
show  that it may lead to cosmological problems associated with
large isocurvature fluctuations and overproduction of dangerous
relics such as gravitinos and moduli fields.  We also note that
the  setting of  initial conditions for the stage of reheating in
these models should be reconsidered. All of these problems can
be resolved in the context of the recently proposed scenario of
instant preheating if there exists an interaction ${g^2}
\phi^2\chi^2$ of the inflaton field $\phi$  with another scalar
field $\chi$. We show that the mechanism of instant preheating in
NO models is much more efficient than the usual mechanism of
gravitational particle production even if the coupling constant
$g^2$ is extremely small, $10^{-14} \ll g^2 \ll 1$.
\end{abstract} \pacs{PACS: 98.80.Cq  \hskip 2.5cm
CITA-99-12~~~~~~SU-ITP-99-13
 \hskip 2.5cm  hep-ph/9903350} \vskip2pc]

\section{Introduction}

Usually it is assumed that the inflaton field $\phi$ after
inflation rolls down to the minimum of its effective potential
$V(\phi)$, oscillates, and eventually decays.  The stage of
oscillations of the inflaton field  is a necessary part of the
standard mechanism of reheating of the universe \cite{DL,KLS}.

However, there exist some models where the inflaton potential
$V(\phi)$ gradually decreases at large $\phi$ and does not have a
minimum. In such theories the inflaton field $\phi$ does not
oscillate after inflation,  so the standard mechanism of reheating
does not work there.

Investigation of  inflationary models of this type has been rather
sporadic \cite{ford,spok,joyce,PV}, and each new author has given
them a new name, such as deflation \cite{spok}, kination
\cite{joyce}, and quintessential inflation \cite{PV}. However, the
universe does not deflate in these models, and in general they are
not related to the theory of quintessence. From our perspective,
the main distinguishing feature of inflationary models of this
type is the non-oscillatory behavior of the inflaton field, which
makes the standard mechanism of reheating inoperative. Therefore
we will call such models ``non-oscillatory models,'' or simply
``NO models.'' In addition to describing the most essential
feature of this class of theories which makes reheating
problematic, this name  reflects the rather negligent attitude
towards these models which existed until now.

 One of the reasons why NO models have not attracted much attention
was the absence of an efficient mechanism of reheating. For a long
time it was believed that the only mechanism of reheating possible
in NO models was the gravitational particle production
\cite{ford,spok,joyce,PV} which occurs because of the changing
metric in the early universe \cite{grib}. This mechanism is very
inefficient, which may lead to certain cosmological problems.

However, recently the situation  changed. The mechanism of instant
preheating which was found in \cite{inst} is very efficient, and
it works in NO models even better than in the models where
$V(\phi)$ has a minimum.

In this paper we will describe various features of NO models.
First of all, we will discuss the problem of initial conditions in
these models, which in our opinion has not been properly addressed
before.  The standard assumption made in \cite{ford,spok,joyce,PV}
is that at the end of inflation in NO models one has a large and
heavy inflaton field $\phi$ which rapidly changes and creates
light particles $\chi$ minimally coupled to gravity from a state
where the classical value of the field $\chi$ vanishes. We will
show that this setting of the problem should be reconsidered. If
the fields $\phi$ and $\chi$ do not interact (which was the
standard assumption of Refs. \cite{ford,spok,joyce,PV}), then at
the end of inflation the field $\chi$ typically does not vanish.
Usually  the last stages of inflation are driven by the light
field $\chi$ rather than by the heavy field $\phi$. But in this
case reheating occurs due to oscillations of the field $\chi$, as
in the usual models of inflation.

In addition to reexamining the problem of initial conditions, we
will point out potential difficulties
associated with isocurvature perturbations and gravitational
production of gravitinos and moduli fields in NO models.

In order to provide a consistent setting for the NO models one
needs to introduce interaction between the fields $\phi$ and
$\chi$. This resolves the problem of initial conditions in these
models and makes it  possible to have a non-oscillatory behavior
of the inflaton field after inflation. We show that all of
these problems can be resolved in the context of the recently
proposed scenario of instant preheating  \cite{inst} if there is
an interaction ${g^2\over 2} \phi^2\chi^2$ of the inflaton field
$\phi$  with another scalar field $\chi$, with $g^2 \gg 10^{-14}$.
In this case the mechanism of instant preheating in NO models is
much more efficient than the usual mechanism of gravitational
particle production studied in \cite{ford,spok,joyce,PV}.

\section{On the initial conditions in NO models without
interactions}\label{init}

 NO models considered in \cite{ford,spok,joyce,PV} described an
inflaton field which does not interact with other fields except
gravitationally. As an example, we will consider here the simplest
theory of the inflaton field  $\phi$ with an effective potential
$V(\phi)$ which behaves as ${\lambda\over 4}\phi^4$ at $\phi < 0$,
and (gradually) vanishes when $\phi $ becomes positive. In
addition, in accordance with \cite{PV},  we will consider a light
scalar field $\chi$ which is not coupled to the inflaton field
$\phi$, and which is minimally coupled to gravity. Reheating in
this model occurs because of the gravitational production of
$\chi$ particles. Application of the general theory of
gravitational particle creation \cite{grib} to the last stages of
inflation and immediate stages after inflation  was considered in
many papers; see in particular \cite{ford,grav}. However, this
theory (and the interpretation of its results)  may change
dramatically if one investigates initial conditions for inflation
and studies quantum fluctuations produced during inflation.

In particular, in all previous works on NO models it was assumed
that at the beginning of inflation $|\phi|$ is very large and
$\chi= 0$. Let us show that in this case at the end of the stage
of inflation driven by the field $\phi$ the long-wavelength
fluctuations of the field $\chi$ typically become so large that it
leads to a new stage of inflation which is driven not by the field
$\phi$, but by the field $\chi$. This conclusion is rather general
and may be extended to other models of $V(\phi)$. The explanation
goes back to the paper \cite{KL87}, where it was found that in the
presence of several scalar fields the last stage of multiple
inflation is typically driven by the lightest scalar field.

Indeed, the field $\phi$ during inflation obeys the following
equation:
\begin{equation}\label{1a}
3H\dot\phi= - \lambda_\phi \phi^3  .
\end{equation}
Here
\begin{equation}\label{2a}
H= \sqrt{2\pi\lambda_\phi\over 3}~ {\phi^2\over M_p}  .
\end{equation}
These two equations yield the solution \cite{chaot}
\begin{equation}\label{3a}
\phi =  \phi_0 ~\exp\left(-\sqrt{\lambda_\phi\over 6 \pi} M_p
t\right).
\end{equation}
 If the field $\chi$ is very light, then in each time interval
$H^{-1}$ during inflation fluctuations $\delta \chi = {H\over
2\pi}$ will be produced. The equation describing the growth of
fluctuations of the field $\chi$ can be written as follows:
\begin{equation}\label{4a}
{d\langle\chi^2 \rangle\over dt}  = {H^3\over 4\pi^2} .
\end{equation}

In de Sitter space with $H = const$ this equation would give
\cite{book}
\begin{equation}\label{5aa}
{ \langle\chi^2 \rangle }  = {H^3\, t\over 4\pi^2}.
\end{equation}

For the theory under consideration $H$ depends on time, and   Eq.
(\ref{4a}) reads
\begin{equation}\label{5a}
{d\langle\chi^2 \rangle\over dt}  = {\lambda_\phi \sqrt
\lambda_\phi\over 3\sqrt{6\pi}} {\phi_0^6 \over M_p^3}
{}~\exp\left(-\sqrt{6\lambda_\phi\over \pi} M_p t\right).
\end{equation}

 The result of
integration at large $t$ converges to
\begin{equation}\label{6a}
{ \langle\chi^2 \rangle }  =
  {\lambda_\phi\,  \phi_0^6\over 18\, M_p^4} .
\end{equation}

These fluctuations from the point of view of a local observer look
like a classical scalar field $\chi$ which is homogeneous on the
scale $H^{-1}$ and which has a typical amplitude
\begin{equation}\label{7a}
\bar \chi = \sqrt{\langle\chi^2 \rangle}  = \sqrt{\lambda_\phi
\over 18}~{\phi_0^3\over M_p^2} .
\end{equation}

This quantity is greater than $M_p$ for
\begin{equation}\label{8a}
\phi_0 \gtrsim \lambda_\phi^{-1/6} M_p  .
\end{equation}
This condition is quite natural. For example, if, in accordance
with \cite{chaot}, inflation begins at $V \sim M_p^4$, one has
$\phi_0 \sim \lambda_\phi^{-1/4} M_p $, which is much greater than
$\lambda_\phi^{-1/6} M_p$.

If the field $\chi$ has a  shallow polynomial potential such as
$m^2_\chi \chi^2/2$  (with a small mass $m_\chi$), or
$\lambda_\chi \chi^4/4$ (with small $\lambda_\chi$), then the
existence of a homogeneous field $\bar \chi \gtrsim M_p$ leads to
a new stage of inflation. This stage will be driven by the light
field $\chi$, and it  will begin after the end of the stage of
inflation driven by the field $\phi$.

 The condition (\ref{8a})  coincides
with the condition that chaotic inflation with respect to the
field $\phi$ enters the stage of self-reproduction \cite{Eternal}.
In this regime the field $\phi$ may stay large even much longer
than is suggested by our classical equations which do not take
self-reproduction into account. As a result, fluctuations of the
field $\chi$ will be even greater, and the last stage of inflation
will be driven not by the field $\phi$ but by the lighter field
$\chi$.

 In such a case, the results of
gravitational particle production obtained in all   previous
papers on NO models do not apply. Instead of particles $\chi$
produced at the end of inflation driven by the field $\phi$ (or in
addition to these particles), we have long-wavelength fluctuations
of the field $\chi$ which initiate a new stage of inflation driven
by the field $\chi$.

A similar result can be obtained in the model $V(\phi)  = {m^2
\over 2} \phi^2$. In this case after inflation driven by the field
$\phi$ one has $\bar \chi = \sqrt{ \langle\chi^2 \rangle }  =   {m
\phi_0^2/(\sqrt 3 M_p)}$. This leads to inflation with respect to
the field $\chi$ (i.e. one has $\bar \chi > M_p$) for
 $\phi_0 \gtrsim M_p \sqrt {M_p\over m}$. Again, according to
\cite{chaot}, the most natural initial value of $\phi$ is $\phi_0
\sim {M_p^2 \over m} \gg M_p \sqrt {M_p\over m}$.

This suggests that if the field $\chi$ is very light (which was
assumed in \cite{spok,joyce,PV}), then it is this field rather
than the field $\phi$ that is responsible for the end of
inflation. Therefore instead of studying gravitational production
of the field $\chi$ due to the non-oscillatory motion of the field
$\phi$ during preheating in this NO model, one should study the
  mechanism
of production  of particles of the field $\phi$ by the oscillating
field $\chi$.

 Of course, one can avoid some of these problems by choosing a specific
set of scalar fields and a specific version  of the theory   which
does not allow any of these scalar fields except the field $\phi$
to drive inflation.   For example, if the field $\chi$ is an axion
field (and no other light scalar fields are present), then it
simply cannot be  large enough to be responsible for inflation.
One can also assume that the field $\chi$ is nonminimally coupled
to gravity and has a large effective mass $O(H)$ during inflation
(see Sect. IV). Then the long-wavelength fluctuations of this
field will not be produced. Thus there are some ways to overcome
the problem mentioned above. But in general  this problem is very
serious, and one should be aware of its existence.

\section{Isocurvature perturbations in the NO  models}

In the previous section we showed that even if one assumes that
$\chi = 0$ at the beginning of inflation, the assumption that one
still has $\chi=0$  at the beginning of reheating in general is
incorrect. The long-wavelength perturbations of the field $\chi$
generated during inflation typically are very large, and they look
like a large homogeneous classical field $\chi$.

Now we will consider a more general question: If the two fields
$\phi$ and $\chi$ do not interact, then why should we assume that
one of them should vanish in the beginning of inflation? And if it
does not vanish, then how does it change the whole picture?

Suppose for example that the field $\chi$ is a Higgs field
\cite{PV} with a relatively small mass and with a large coupling
constant $\lambda_\chi \gg \lambda_\phi $. The total effective
potential in this theory (for $\phi < 0$) is given by
\begin{equation}\label{9a}
V(\phi,\chi) =  {\lambda_\phi \over 4} \phi^4+
  {\lambda_\chi \over 4}(\chi^2-v^2)^2 .
\end{equation}
Here $v$ is the amplitude of spontaneous symmetry breaking, $v \ll
M_p$. During inflation and at the first stages of reheating this
term can be neglected, so we will study the simplified model
\begin{equation}\label{10a}
V(\phi,\chi) =  {\lambda_\phi \over 4} \phi^4+
  {\lambda_\chi \over 4}\chi^4 .
\end{equation}

This model was first analyzed in \cite{KL87}. It is directly
related to the Peebles-Vilenkin model \cite{PV} if the field
$\chi$ is the Higgs boson field with a small mass $m$. This model
exhibits the following unusual feature.

In general, at the beginning of inflation one has both $\phi \not
= 0$ and $\chi \not = 0$. Thus, unlike in the previous subsection,
we will not assume that $\chi = 0$, and instead of studying
quantum fluctuations of this field which can make it large, we
will assume that it could be large from the very beginning.

Even though the fields  $\phi$ and $\chi$  do not interact with
each other directly, they move towards the state $\phi = 0$ and
$\chi = 0$ in a coherent way. The reason is that the motion of
these fields is determined by the same value of the Hubble
constant $H$.

The equations of motion for both fields during inflation look as
follows:
\begin{equation}\label{11a}
3H\dot\phi= - \lambda_\phi \phi^3  .
\end{equation}
\begin{equation}\label{12a}
3H\dot\chi= - \lambda_\chi \chi^3  .
\end{equation}
These equations imply that
\begin{equation}\label{13a}
{d\phi\over \lambda_\phi \phi^3}={d\chi\over \lambda_\chi \chi^3}
,
\end{equation}
which yields the general solution
\begin{equation}\label{14a}
{1\over \lambda_\phi \phi^2}={1\over \lambda_\chi \chi^2} +
{1\over \lambda_\phi \phi_0^2}-{1\over \lambda_\chi \chi_0^2} ,
\end{equation}
Since the initial values of these fields are much greater than the
final values, at the last stages of inflation one has  \cite{KL87}
\begin{equation}\label{15a}
{ \phi\over \chi}=\sqrt{\lambda_\chi \over \lambda_\phi} .
\end{equation}

 Suppose $\lambda_\phi \ll \lambda_\chi$. In this case the
``heavy'' field $\chi$ rapidly rolls down, and then from the last
equation it follows, rather paradoxically, that the Hubble
constant at the end of inflation is dominated  by the ``light''
field $\phi$.  Thus we can consistently consider the creation of
fluctuations of the field $\chi$ ($\chi$ particles) at the end of
and after the last inflationary stage driven by the $\phi$ field.
But now these fluctuations occur on top of a nonvanishing
classical field $\chi$.

To study the behavior of the classical fields $\phi$ and $\chi$
and their fluctuations analytically, one should remember that
during the inflationary stage driven by $\phi$ one has $H =
\sqrt{2\lambda_\phi\pi\over 3}~{\phi^2\over M_p}$, as in the
previous  section. In this case, as before, the solution for the
equation of motion of the field $\phi$ is given by \cite{chaot}
\begin{equation}\label{16a}
  \phi =\phi_0~ \exp\left(-\sqrt{\lambda_\phi\over 6\pi}M_pt\right) .
\end{equation}
Meanwhile, according to Eq. (\ref{15a}),
\begin{equation}\label{17a}
  \chi =\phi_0\sqrt{\lambda_\phi \over \lambda_\chi}~
\exp\left(-\sqrt{\lambda_\phi\over 6\pi}M_pt\right) ,
\end{equation}
 whereas for perturbations of the field $\chi$  one has:
\begin{equation}\label{17aa}
 \delta \chi =\delta\chi_0 ~
\exp\left(-3\sqrt{\lambda_\phi\over 6\pi}M_pt\right) .
\end{equation}

 Let us consider, for example, the behavior of the fields and their
fluctuations at the end of inflation, starting from the moment
$\phi = \phi_i$. One may take, for example, $\phi_i\sim 4 M_p$,
which corresponds to a point approximately 60 e-folds before the
end of inflation. The fluctuations $\delta\chi_i \sim
H(\phi_i)/2\pi$  decrease according to (\ref{17aa}), and at the
end of inflation one gets
\begin{equation}\label{17aaa}
{ \delta \chi \over \chi}={H(\phi_i)\over 2\pi  \chi_i}
\exp\left(-2\sqrt{\lambda_\phi\over 6\pi}M_pt\right) =  {\sqrt {
\lambda_\chi} \phi_i
  \over  \sqrt { 6 \pi}  M_p} {\phi_e^2\over \phi_i^2} .
\end{equation}

 Here $\phi_e \sim 0.3 M_p$ corresponds to the end of  
inflation.\footnote{We are grateful to Peebles and
Vilenkin for pointing out that the factor ${\phi_e^2\over
\phi_i^2}$ should be present in this equation.}
After that moment the fields $\phi$ and $\chi$ begin oscillating,
and the ratio  of $ \delta \chi$ to the amplitude of oscillations
of the field $\chi$ remains approximately constant.  This gives the  
following
estimate for the amplitude of
isocurvature perturbations in this model:
\begin{equation}\label{19a}
{\delta V(\chi)\over V(\chi) } \sim 4 { \delta \chi\over \chi} =
2\times 10^{-2} \sqrt {\lambda_\chi}  \ .
\end{equation}

Initially  perturbations of $V(\chi)$ give a negligibly small
contribution to perturbations of the metric because $V(\chi) \ll
V(\phi)$; that is why they are called isocurvature perturbations.
However, the main idea of preheating in NO models is that
eventually $\chi$ fields or the products of their decay  will give
the dominant contribution to the energy-momentum tensor because
the energy density of the field $\phi$ rapidly vanishes  due to
the expansion of the universe ($\rho_\phi \sim a^{-6}$). However,
because of the inhomogeneity of the distribution of the field
$\chi$ (which will be imprinted in the density distribution of the
products of its decay on scales greater than $H^{-1}$), the period
of the dominance of matter over the scalar field $\phi$ will
happen at different times in different parts of the universe. In
other words, the epoch when the universe  begins expanding as $a
\sim \sqrt t$ or $a \sim  t^{2/3}$ instead of $a \sim t^{1/3}$
will begin at different moments $t$ (at different total densities)
in different parts of the universe. Starting from this time the
isocurvature fluctuations (\ref{19a}) will produce metric
perturbations, and, as a result, perturbations of CMB radiation.

Note that if the equation of state of the field $\chi$ or of the
products of its decay coincided with the equation of state of the
scalar field $\phi$ after inflation,
  fluctuations of the field $\chi$ would not induce  any anisotropy of CMB
radiation. For example, these fluctuations would be harmless if
the field $\chi$ decayed into ultrarelativistic particles with the
equation of state $p = \rho/3$ and if the equation of state of the
field $\phi$ at that time were also given by $p = \rho/3$.
However, in our case the field $\phi$ has equation of state $p =
\rho$, which is quite different from the equation of state of the
field $\chi$ or of its decay products.

 Isocurvature fluctuations lead to approximately 6 times greater
 large scale anisotropy of
the cosmic microwave radiation as compared with adiabatic
perturbations. To avoid cosmological problems, one would need to
have ${\delta V(\chi)\over V(\chi) } \lesssim 5\times 10^{-6}$. If
$\chi$ is the Higgs field with $\lambda_\chi \gtrsim  10^{-7}$,
then the  perturbations discussed above will be unacceptably
large. This may be a rather serious problem. Indeed, one may
expect to have many scalar fields in realistic theories of
elementary particles. To avoid large isocurvature fluctuations
each of these fields must be extremely weakly coupled, with
$\lambda_\chi \lesssim  10^{-7}$,

The general conclusion is that the theory of reheating in NO
models, as well as their consequences for the creation of the
large-scale structure of the universe, may be quite different from
what was anticipated in the first papers on this subject. In the
simplest versions of such models inflation typically does not end
in the state $\chi = 0$, and large isocurvature fluctuations are
produced.

\section{Cosmological production of gravitinos and moduli fields}

If the inflaton field $\phi$ is sterile, not interacting with any
 other fields, the elementary particles constituting the universe
should be produced gravitationally due to the variation of the
 scale factor $a(t)$ with time. This was one of the basic
assumptions of all papers on NO models  \cite{ford,spok,joyce,PV}.
Not all species can be produced this way, but only those which are
not conformally invariant. Indeed, the metric of the Friedmann
universe is conformally flat. If one considers, for example,
massless scalar  particles $\chi$  with an additional term
$-{1\over 12} \chi^2 R$ in  the Lagrangian (conformal coupling),
one can make conformal transformations of $\chi$ simultaneously
with transformations of the metric and find that the theory of
$\chi$ particles in the Friedmann universe is equivalent to their
theory in flat space. That is why such particles would not be
created in an expanding universe.

 Since conformal coupling is a rather special
requirement, one expects a number of different species to be
produced. An apparent advantage of gravitational particle
production is its universality  \cite{grav}.
 There is a kind of ``democracy" rule for all
particles non-conformally coupled to gravity: the density of such
particles produced at the end of inflation is $\rho_X \sim
\alpha_X H^4$, where $\alpha_X \sim 10^{-2}$ is a numerical factor
specific for different species and $H$ is the Hubble parameter at
the end of inflation.

Unfortunately, democracy does not always work; there may be too
many dangerous relics produced by this universal mechanism. One of
the potential problems is related to the
 overproduction of gravitons mentioned in \cite{PV}. In order to
solve it one needs to have models with a very large number of
types of light particles. This is difficult but not impossible
\cite{PV}. However, even more difficult problems will arise if NO
models are implemented in supersymmetric theories of elementary
particles.

For example, in supersymmetric theories one may encounter many
flat directions of the effective potential associated with moduli
fields. These fields usually are very stable. Moduli particles
decay very late, so in order to avoid cosmological problems the
energy density of the moduli fields must be many orders of
magnitude smaller than the energy density of other particles
\cite{modmass,constraint}.

Moduli fields typically are not conformally invariant. There are
several different effects which add up to give them  masses $C H$
during expansion of the universe, with $C = O(1)$ (in general, $C
$ is not a constant) \cite{modmass}. This is very similar to what
happens if, for example, one adds a term $-{\xi\over 2}  R\phi^2$
to the lagrangian of a scalar field.  Indeed, during inflation $R
= 12 H^2$, so this term leads to the appearance of a contribution
to the  mass of the scalar field $\Delta m^2 = 12 \xi H^2$.
Conformal coupling would correspond to $m^2 = 2 H^2$.

According to \cite{ford}, the energy density of scalar particles
produced gravitationally at the end of inflation is given by
$10^{-2} H^4  (1 - 6\xi)^2$. Thus, unless the constant $C$ is
fine-tuned to mimic conformal coupling,   we expect that in
addition to the energy of classical oscillating moduli fields, at
the end of inflation one has gravitational production of moduli
particles with energy density $\sim 10^{-2}   H^4$, just as for
all other conformally noninvariant particles.

 In usual inflationary models one also encounters the moduli problem
if the energy of classical oscillating moduli fields is too large
\cite{modmass}. Here we are discussing an independent problem
which appears even if there are no classical oscillating moduli.
Indeed, in NO models all particles created by gravitational
effects at the end of  inflation will have similar energy density
$\sim 10^{-2}   H^4$. But if the energy density of moduli fields
is not extremely strongly suppressed as compared with the energy
density of other particles, then such models will be ruled out
\cite{modmass,constraint}.

A similar problem appears if one considers the possibility of
gravitational (nonthermal) production of gravitinos.  Usually it
is assumed that gravitinos have mass $m_{3/2}=10^2 - 10^3$ GeV,
which is much smaller than the typical value of the Hubble
constant at the end of inflation. Therefore naively one could
expect that gravitinos, just like massless fermions of spin $1/2$,
are (almost exactly) conformally invariant and should not be
produced due to expansion of the Friedmann universe.

However, in the framework of supergravity, the background metric
is generated by inflaton field(s) $\phi_j$ with an effective
potential constructed from the superpotential $W(\phi_j)$. The
gravitino mass in the early universe acquires a contribution
proportional to $W(\phi_j)$. Depending on the model, the gravitino
mass soon after the end of inflation may be of the same order as
$H$ or somewhat smaller, but typically it is much greater than its
present value $m_{3/2}$.

A general investigation of the behavior of gravitinos in the
Friedmann universe shows that the gravitino field in a
self-consistent Friedmann background supported by scalar fields is
not conformally invariant \cite{KKLV}.  For example, the effective
potential $\lambda\phi^4$ can be obtained from the superpotential
$\sqrt\lambda\phi^3$ in the global supersymmetry limit. This leads
to a gravitino mass $\sim \sqrt\lambda\phi^3/M_p^2$. At the end of
inflation  $\phi \sim M_p$, and therefore the gravitino mass is
comparable to the Hubble constant $H \sim \sqrt\lambda\phi^2/M_p$.
This implies strong breaking of conformal invariance.

The theory of gravitational production of gravitinos is strongly
model-dependent, and in some models it might be possible to
achieve a certain suppression of their production as compared to
the production of other particles. The problem is that, just like
in the situation with the moduli fields, this suppression must be
extraordinary strong. Indeed, to avoid cosmological problems one
should suppress the number of gravitinos as compared to the number
of other particles by a factor of about $10^{-15}$
\cite{gravitinos}. We will present a more detailed discussion of
the cosmological production of gravitinos and moduli in a separate
publication \cite{KKLV}.

 The gravitino/moduli problem and the problem of isocurvature
perturbations are interrelated in a rather nontrivial way. Indeed,
the gravitino and moduli problems are especially severe if the
density of gravitinos and/or moduli particles produced during
 reheating is of the same order of magnitude as the energy
density of scalar fields $\chi$. We assumed, according to
\cite{ford,PV}, that the energy density of the fields $\chi$ after
inflation is $O(10^{-2} H^4)$. But this statement is not always
correct. It was derived in \cite{ford} under an assumption that
particle production occurs during a short time interval when the
equation of state changes. Meanwhile in inflationary cosmology the
long-wavelength fluctuations of the field $\chi$ minimally coupled
to gravity are produced during inflation all the time when the
Hubble constant $H(t)$ is smaller than the mass of the $\chi$
particles $m_\chi$. The energy density of $\chi$ particles
produced during inflation will contain a contribution $\rho_0 =
{m^2_\chi\over 2} \langle\chi^2\rangle$, which may be many orders
of magnitude greater than $10^{-2} H^4$.

For the sake of argument, one may consider inflation in the theory
${\lambda\over 4}\phi^4$ and take $m_\chi$ equal to the value of
$H$ at the end of inflation, $m_\chi \sim \sqrt\lambda_\phi M_p$.
Then, according to Eq. (\ref{6a}), after inflation one has
$\rho_0 \sim  {\lambda_\phi\, m^2_\chi \phi_0^6\over 36\, M_p^4}
\sim 10^{-2} H^4  \left({\phi_0 \over M_p}\right)^6
 \gg 10^{-2} H^4$, because $\phi_0 \gg M_p$.

This is the same effect which we discussed in
Section~{\ref{init}}: If $\phi_0$ is large enough, we may even
have a second stage of inflation driven by the large energy
density of the fluctuations of the field $\chi$. But even if
$\phi_0$ is not large enough to initiate the second stage of
inflation, it still must be much greater than $M_p$ to drive the
first stage of inflation, which makes the standard estimate $\rho
\sim 10^{-2} H^4$ incorrect \cite{more}.

 There is one more effect which should be considered, in
addition to gravitational particle production. The effective mass
of the particles $\phi$ at $\phi < 0$ is given by $\sqrt{3\lambda}
\phi$. At the end of inflation, at $\phi \sim M_p$, this mass is
of the same order as the Hubble constant $\sim \sqrt{ \lambda}
\phi^2/M_p$. Then, within the Hubble time $H^{-1}$ the field
$\phi$ rolls to the valley at $\phi > 0$ and its mass vanishes.
This is a non-adiabatic process; the mass of the scalar field
changes by $O(H)$ during the time $O(H^{-1})$. As a result, in
addition to gravitational particle production there is an equally
strong production of particles $\phi$ due to the nonadiabatic
change of their mass \cite{more,f2}.

This may imply that the fraction of energy in gravitinos will be
much smaller than previously expected, simply because the fraction
of energy in the fluctuations of the field $\chi$ will be much
larger. But there is no free lunch.  For example, the production
of large number of nearly massless particles $\phi$ may lead to
problems with nucleosynthesis. Large inflationary fluctuations of
the field $\chi$ can create large isocurvature fluctuations. In
the end of Section~{\ref{init}} we mentioned that one can avoid
this problem if one assumes, for example,   that the fields $\chi$
acquire effective mass $O(H)$ in an expanding universe. Then their
fluctuations will not be produced during inflation. But in such a
case their density after inflation will be given by $10^{-2} H^4$,
and therefore we do not have any relaxation of the gravitino and
the moduli problems.

\section{Saving NO models: Instant preheating}

As we will see, the problems discussed above will not appear in
theories of a more general class, where the fields $\phi$ and
$\chi$ can interact with each other. We will consider a model with
the interaction ${g^2\over 2}\phi^2\chi^2$. First we will show
that in this case it really makes sense to study preheating
assuming that $\chi = 0$. Then we will describe the scenario of
instant preheating, which allows a very efficient energy transfer
from the inflaton field to particles $\chi$.

\subsection{Initial conditions for inflation and reheating in the
model with interaction ${g^2\over 2}\phi^2\chi^2$}

Consider a theory with an effective potential dominated by the
term $V(\phi,\chi) = {g^2\over 2} \phi^2\chi^2$.  This means that
we will assume that the constant $g$ is large enough for us to
temporarily neglect the terms ${\lambda_\phi \over 4} \phi^4+
  {\lambda_\chi \over 4}\chi^4$ in the discussion of initial conditions.

In this case the Planck boundary is given by the condition
\begin{equation}\label{1}
{g^2\over 2} \phi^2\chi^2 \sim M_p^4 \ ,
\end{equation}
which defines a set of four hyperbolas
\begin{equation}\label{2}
 g |\phi ||\chi| \sim M_p^2 \ .
\end{equation}
At larger values of $\phi$ and $\chi$ the density is greater than
the Planck density, so the standard classical description of
space-time is impossible there. On the other hand, the effective
masses of the fields should be smaller than $M_p$, and
consequently the curvature of the effective potential cannot be
greater than $M_p^2$. This leads to two additional conditions:
\begin{equation}\label{3}
  |\phi | \lesssim g^{-1} M_p, ~~~~~~|\chi| \lesssim g^{-1} M_p.
\end{equation}
We assume that $g \ll 1$. Suppose for definiteness that initially
the fields $\phi$ and $\chi$ belong to the Planck boundary
(\ref{2}) and that $|\phi |$ is half-way towards its upper bound
(\ref{3}): $|\phi | \sim g^{-1} M_p/2$. The choice of the
coefficient $1/2$ here is not essential; we only want to make sure
that the field $\chi$ initially is of order $M_p$, but it can be
slightly greater than $M_p$, This allows for an extremely short
stage of inflation when the field $\chi$ rolls down towards $\chi
= 0$.

The equations for the two fields are
\begin{equation}\label{4}
 \ddot \phi + 3H \dot \phi = -{g^2 } \phi \chi^2.
\end{equation}
and
\begin{equation}\label{5}
 \ddot \chi + 3H \dot \chi = -{g^2 } \phi^2 \chi .
\end{equation}
The curvature of the effective potential in the $\phi$ direction
initially is   $\sim g^2\chi^2 \sim g^2 M_p^2$, which is very
small compared to the initial value of   $H^2 \sim M_p^2$. Thus
the field $\phi$ will move very slowly, so one can neglect  the
term $\ddot \phi$ in Eq. (\ref{4}).
\begin{equation}\label{6}
 3H \dot \phi = -{g^2 } \phi \chi^2.
\end{equation}
If the field $\phi$ changes slowly, then the field   $\chi$
behaves as in the theory ${m^2_\chi\over 2} \chi^2$ with $m_\chi
\sim g|\phi|$ being slightly smaller than $M_p$ and with the
initial value of $\chi$ being slightly greater than $M_p$. This
leads to a very short stage of inflation which ends within a few
Planck times. After this short stage the field $\chi$ rapidly
oscillates.  During this stage the energy density of the
oscillating field drops down as $a^{-3}$, the universe expands as
$a \sim t^{2/3}$, and $H = {2\over 3t}$. Thus the square of the
amplitude of the oscillations of the field $\chi$ decreases as
follows: $\chi^2 \sim \chi_0^2 a^{-3} \sim t^{-2}$. This leads to
the following equation for the field $\phi$:
\begin{equation}\label{8}
 { \dot \phi\over \phi } \sim -{g^2  \over t }.
\end{equation}
The solution of this equation is ${ \phi } = \phi_0 \left({t
\over t_0 }\right)^{-g^{2}}$ with $t_0 \sim  M_p^{-1}$,   and
$\phi_0 \sim -M_p/g$. (The condition $t_0 \sim  M_p^{-1}$ follows
from the fact that the initial value of $H = {2\over 3t}$ is not
much below $M_p$.) This gives
\begin{equation}\label{9}
 { \phi }  \sim -{M_p\over
 g} ~\left(M_pt\right)^{-g^2} .
\end{equation}
The inflaton field $\phi$ becomes equal to   $-M_p$ after the
exponentially large time
\begin{equation}\label{10}
t \sim M_p^{-1} \left({1\over g }\right)^{g^{-2}}.
\end{equation}
During this time the energy of oscillations of the field $\chi$
becomes exponentially small, and the small term
${\lambda_\phi\over 4} \phi^4$ which we neglected until now
becomes the leading term driving the scalar field $\phi$. At this
stage we will have the usual chaotic inflation scenario with
$|\phi| > M_p$ and with the fields evolving along the direction
$\chi = 0$.

 Thus in the presence of the
interaction term ${g^2\over 2}\phi^2\chi^2$ one can indeed
consider inflation and reheating with $\chi = 0$. As we have seen,
this possibility was rather problematic in the models where $\phi$
and $\chi$ interacted only gravitationally.

The effective mass of the field $\chi$ during inflation is
$g|\phi|$, which is much greater than the Hubble constant $\sim
{\sqrt\lambda\phi^2\over M_p}$ for ${g^2\over \lambda} \gg
{\phi^2\over M_p^2}$. In realistic versions of this model one has
$\lambda \sim 10^{-13}$ \cite{book}, and  $g^2 \gg \lambda$.
Therefore long-wavelength fluctuations of the field $\chi$ are not
produced during the last stages of inflation, when $\phi \sim
M_p$.

A similar conclusion is valid if at the last stages of inflation
  the effective potential of the field $\phi$ is quadratic,
$V(\phi) = {m^2\over 2}\phi^2$. In this case $H \sim {m\phi\over
M_p}$, and  inflationary fluctuations of the field $\chi$ are not
produced for $g \gg {m\over M_p}$. In realistic versions of this
model one has $m \sim 10^{-6} M_p$ \cite{book}, and fluctuations
$\delta \chi$ are not produced if $g^2 \gg 10^{-12}$. This means
that the problem of isocurvature fluctuations does not appear.

\subsection{Instant preheating in NO models}

To explain the main idea of the instant preheating scenario in NO
models,   we will assume for simplicity that $V(\phi) = {m^2\over
2} \phi^2$ for $\phi < 0$, and that $V(\phi)$ vanishes for $\phi >
0$. We will discuss a more general situation later. We will assume
that the effective potential contains the interaction term   ${g^2
\over 2}\phi^2\chi^2$, and that $\chi$ particles have the usual
Yukawa interaction $h \bar\psi\psi \chi$ with fermions $\psi$. For
simplicity, we will assume here that $\chi$ particles do not have
any bare mass, so that their effective mass is equal to $g|\phi|$.

In this model inflation ends when the field $\phi$ rolls from
large negative values down to  $\phi \sim -0.3 M_p$ \cite{book}.
Production of particles $\chi$ begins when the effective mass of
the field $\chi$ starts to change nonadiabatically, $|\dot m_\chi|
\gtrsim m^2_\chi$, i.e. when $g|\dot \phi|$ becomes greater than
$g^2\phi^2$.

This happens only when the field $\phi$ rolls close to $\phi = 0$,
and the velocity of the field is   $|\dot\phi_0| \approx m M_p/10
\approx 10^{-7} M_p$  \cite{inst}. (In the theory ${\lambda\over
4}\phi^4$  with $\lambda = 10^{-13}$ one has a somewhat smaller
value $|\dot\phi_0| \approx 6\times 10^{-9} M_p^2$.) The process
becomes nonadiabatic for $g^2 \phi^2 \lesssim g|\dot\phi_0|$, i.e.
for $ -\phi_* \lesssim  \phi \lesssim \phi_*$, where $\phi_* \sim
\sqrt{|\dot\phi_0|\over g}$ \cite{KLS}.  Note that for $g\gg
10^{-5}$   the interval  $ -\phi_* \lesssim  \phi \lesssim \phi_*$
is very narrow: $\phi_* \ll M_p/10$. As a result, the process of
particle production occurs nearly instantaneously, within the time
\begin{equation}\label{time}
\Delta t_* \sim  {\phi_*\over |\dot\phi_0|} \sim (g
|\dot\phi_0|)^{-1/2} .
\end{equation}
This time interval is much smaller than the age of the universe,
so all effects related to the expansion of the universe can be
neglected during the process of particle production. The
uncertainty principle implies in this case that the created
particles will have typical momenta $k \sim   (\Delta t_*)^{-1}
\sim  (g |\dot\phi_0|)^{1/2}$. The occupation number  $n_k$ of
$\chi$ particles with momentum $k$ is equal to zero
 all the time when it moves toward $\phi = 0$. When it reaches $\phi
= 0$ (or, more exactly, after it moves through the small region  $
-\phi_* \lesssim  \phi \lesssim \phi_*$)  the occupation number
suddenly (within the time $\Delta t_*$) acquires the value
\cite{KLS}
\begin{equation}\label{number}
n_k = \exp\left(-{\pi k^2 \over  g|\dot\phi_0|}\right)  ,
\end{equation}
and this value does not change until the field $\phi$ rolls to the
point $\phi = 0$ again.

A detailed description of this process including the derivation of
Eq. (\ref{number}) was given   in
  the second paper of Ref.
\cite{KLS}; see in particular Eq. (55) there.
 This  equation  (\ref{number}) can be written in a  more general
form.  For example, if the particles $\chi$ have bare mass
$m_\chi$, this equation can be written as follows \cite{inst}:
\begin{equation}\label{general}
n_k = \exp\left(-{\pi (k^2+ m_\chi^2) \over  g|\dot\phi_0|}\right)
{}.
\end{equation}
 This can be integrated to give the
density of $\chi$ particles
\begin{equation}\label{suppr}
 n_{\chi}   =  {1 \over 2\pi^2  }
\int\limits_0^{\infty} dk\,k^2 n_k =  {({g\dot\phi_0})^{3/2} \over
8\pi^3} \exp
 \left(-{\pi  m_\chi^2  \over  g|\dot\phi_0|}\right)   .
\end{equation}
As we already mentioned, in the theory ${m^2\over 2}\phi^2$  with
$m = 10^{-6} M_p$ one has $|\dot\phi_0| = 10^{-7} M_p^2$
\cite{inst}.  This implies, in particular, that  if one takes $g
\sim 1$, then in the theory ${m^2\over 2}\phi^2$ there is no
exponential suppression of production of $\chi$ particles unless
their mass is greater than $m_\chi \sim 2 \times 10^{15} $ GeV.
This agrees with a similar conclusion obtained in
\cite{KLS,Kolb,heavy,chung}.

Let us now concentrate on the case $ m_\chi^2 \ll g|\dot\phi_0|$,
when the number of produced particles is not exponentially
suppressed. In this case the number density of particles at the
moment of their creation is given by ${({g\dot\phi_0})^{3/2} \over
8\pi^3}$, but then it decreases as $a^{-3}(t)$:
\begin{equation}\label{numberdens}
 n_{\chi}   =   {({g\dot\phi_0})^{3/2} \over 8\pi^3 a^{3}(t)}  \ .
\end{equation}
Here we take $a_0 = 1$ at the moment of particle production.

Particle production occurs only in a small vicinity of $\phi = 0$.
Then the field $\phi$ continues rolling along the flat direction
of the effective potential with $\phi > 0$, and the mass of each
$\chi$ particle grows as $g\phi$.  Therefore the energy density of
produced particles is
\begin{equation}\label{dens}
\rho_{\chi} =   {({g\dot\phi_0})^{3/2} \over 8\pi^3}~{g\phi(t)
\over a^{3}(t)} \ .
\end{equation}
The energy density of the field $\phi$ drops down much faster, as
$a^{-6}(t)$. The reason is that if one neglects backreaction of
produced particles, the energy density of the field $\phi$ at this
stage is entirely concentrated in its kinetic energy density
${1\over 2}\dot\phi^2$, which corresponds to the equation of state
$p = \rho$. We will study this issue now in a more detailed way.

The equation of motion for the inflaton field after particle
production looks as follows:
\begin{equation}\label{11b}
\ddot\phi + 3H\dot  \phi = - g^2\phi \langle \chi^2 \rangle
\end{equation}
We will assume for simplicity that the field $\chi$ does not have
bare mass, i.e. $m_\chi = g\phi$. As soon as the field $\phi$
becomes greater than $\phi^*$ (and this happens practically
instantly, when particle production ends), the particles $\chi$
become nonrelativistic. In this case $\langle \chi^2 \rangle$ can
be easily related to $n_\chi$:
\begin{equation}\label{variance}
\langle \chi^2 \rangle \approx {1\over 2\pi^2}\int{n_k\,k^2 d
k\over \sqrt{k^2+ g^2\phi^2}} \approx  {  n_\chi\over g\phi}
\approx
  {({g\dot\phi_0})^{3/2} \over 8\pi^3 g \phi a^{3}(t)} \ .
\end{equation}
Therefore the equation for the field $\phi$ reads
\begin{equation}\label{roll}
\ddot\phi + 3H\dot  \phi = - g n_\chi = -g {({g\dot\phi_0})^{3/2}
\over 8\pi^3 a^{3}(t)} \ .
\end{equation}
To analyze the solutions of this equation, we will first neglect
backreaction. In this case one has $a \sim t^{1/3}$, $H = {1\over
3t}$, and
\begin{equation}\label{field}
\phi = {M_p\over 2\sqrt {3\pi}}~\log{t\over t_0}\ ,
\end{equation}
where   $t_0 = {1\over 3 H_0} = {M_p\over 2\sqrt {3\pi}\dot\phi_0}
\approx {5\over \sqrt {3\pi}m}$.

 One can easily check that this regime remains intact and backreaction
is unimportant for $ t < t_1 \sim {8 \pi^3\over \sqrt{g^{5
}\dot\phi_0}}$, until the field $\phi$ grows up to
\begin{equation}\label{stop}
\phi_1 \approx {5M_p\over 4\sqrt {3\pi}}~\log{1\over g} \ .
\end{equation}
This equation is valid for $g \ll 1$. For example, for $g =
10^{-3}$ one has $\phi_1 \sim 3M_p$. For $g = 10^{-1}$ one has
$\phi_1 \sim M_p$. Note that the terms in the left hand side of
the Eq. (\ref{roll}) decrease as $t^{-2}$  when the time grows,
whereas the backreaction term  goes as $t^{-1}$.   As soon as the
backreaction becomes important, i.e. as soon as the field  $\phi$
reaches $\phi_1$, it turns back, and returns to $\phi = 0$. When
it reaches $\phi = 0$, the effective potential becomes large, so
the field $\phi$ cannot become negative, and it bounces towards
large $\phi$ again.

Now let us take into account interaction of the $\chi$ field with
fermions. This interaction  leads to decay of the $\chi$ particles
with the decay rate  \cite{KLS}
\begin{equation}\label{7b}
\Gamma( \chi \to \psi \psi ) = { h^2 m_\chi\over 8 \pi} = { h^2 g
|\phi|\over 8 \pi}\ .
\end{equation}
Note that the decay rate grows with the growth of the field
$|\phi|$, so particles tend to decay at  large $\phi$. In our case
the field $\phi$ spends most of the time prior to $t_1$ at $\phi
\sim M_p$ (if it does not decay earlier, see below). The decay
rate at that time is
\begin{equation}\label{7bb}
\Gamma( \chi \to \psi \psi ) \sim { h^2 g M_p\over 8 \pi}\ .
\end{equation}
If $\Gamma_\chi^{-1} < t_1 \sim {8 \pi^3\over g^{5/2}\dot\phi_0}$,
then particles $\chi$ will  decay to fermions $\psi$ at $t < t_1$
and the force driving the field $\phi$ back to $\phi = 0$ will
disappear before the field $\phi$ turns back. In this case the
field $\phi$ will continue to grow, and  its energy density will
continue decreasing anomalously fast, as  $a^{-6}$. This happens
if
\begin{equation}\label{7bbb}
{ h^2 g M_p\over 8 \pi} \gtrsim { g^{5/2}\dot\phi_0^{1/2}\over
8\pi^3} .
\end{equation}
Taking into account that in our case  $\dot\phi_0 \sim {mM_p\over
10}$ and $m \sim 10^{-6} M_p$, one finds that this condition is
satisfied if   $h \gtrsim  5 \times 10^{-3} g^{3/4}$. This is a
very mild condition. For example, it is satisfied for $h > 5
\times 10^{-3}$ if $g = 1$, and for $h > 5 \times 10^{-7}$ if $g =
10^{-4}$.

This scenario is always 100\% efficient. The initial fraction of
energy transferred to matter at the moment of $\chi$ particle
production is not very large, about $10^{-2} g^2$ of the energy of
the inflaton field \cite{inst}. However, because of  the
subsequent growth of this energy due to the growth of the field
$\phi$,  and because of the rapid decrease of kinetic energy of
the inflaton field, the energy density of the $\chi$ particles and
of the products of their decay soon becomes dominant. This should
be contrasted with the usual situation in the theories where
$V(\phi)$ has a minimum. As was emphasized in \cite{KLS},
efficient preheating is possible only in a   subclass of such
models. In many models where $V(\phi)$ has a minimum the decay of
the inflaton field is incomplete, and it accumulates an
unacceptably large energy density compared with the energy density
of the thermalized component of matter. The possibility of having
a very efficient reheating in NO models   may have significant
consequences for inflationary model building.

It is instructive to compare the density of particles produced by
this mechanism  to the density of particles created during
gravitational particle production, which is given by $\rho_\chi
\sim 10^{-2}  H^4  \sim    \rho_\phi {\rho_\phi\over M_p^4}$,
where $\rho_\phi$ is the energy density of the field $\phi$ at the
end of inflation. In the model ${\lambda_\phi\over 4} \phi^4$ one
has $\rho_\phi \sim 10^{-16} M_p^4$, and, consequently,
$\rho_{\chi} \sim \rho_\phi {\rho_\phi\over M_p^4} \sim 10^{-16}
\rho_\phi$. Meanwhile, as we just mentioned,  at the first moment
after particle production in our scenario the energy density of
produced particles is of the order of $10^{-2} g^2 \rho_\phi$
\cite{inst}, and then it grows together with the field  $\phi$
because of the growth of the mass $g \phi$ of each $\chi$
particle. Thus, for $g^2 \gtrsim 10^{-14}$ the number of particles
produced during instant preheating  is much greater than the
number of particles produced by gravitational effects. Therefore
one may argue that reheating of the universe in NO models should
be described using the instant preheating scenario. Typically it
is much more efficient than gravitational particle production.
This means, in particular, that production of normal particles
will be much more efficient than the production of gravitinos and
moduli.

In order to avoid the gravitino problem altogether one may
consider versions of NO models where the particles produced during
preheating remain nonrelativistic for a while. Then the energy
density of gravitinos during this epoch decreases much faster than
the energy density of usual particles. New gravitinos will not be
produced if  the resulting temperature of reheating is
sufficiently small.

\section{Other versions of NO models}

 The
mechanism of particle production described above can work in a
broad class of theories. In particular, since  parametric
amplification of particle production is not important in the
context of the instant preheating scenario, it will work equally
well if the inflaton field couples not to bosons $\chi$ but to
fermions \cite{baacke,GK98}. Indeed, the
  creation
of fermions with mass $g|\phi|$  also occurs because of the
nonadiabaticity of the change of their mass at   $ \phi  = 0$. The
theory of this effect   is very similar to  the theory of the
creation of $\chi$ particles described above; see in this respect
\cite{GK98}.

Returning to our scenario, production of particles $\chi$ depends
on the  interactions between the fields $\phi$ and $\chi$. For
example, one can consider models with the interaction ${g^2\over
2}\chi^2(\phi + v)^2$. Such interaction terms appear, for example,
in supersymmetric models with superpotentials of the type  $W = g
\chi^2(\phi + v)$ \cite{berera}. In such models the mass $m_\chi$
vanishes not at $\phi_1 = 0$, but at $\phi_1 = -v$, where $v$ can
take any value. Correspondingly,  the production of $\chi$
particles occurs not at $\phi = 0$ but at $\phi = -v$. When the
inflaton field reaches  $\phi = 0$,  one has $m_\chi \sim g v$,
which may be very large. If one takes $v \sim M_p$, one can get
$m_\chi \sim g M_p$, which may be as great as $10^{18}$ GeV for $g
\sim 10^{-1}$, or even $ 10^{19}$ GeV for $g \sim 1$. If one takes
$v \gg M_p$, the density of $\chi$ particles produced by this
mechanism will be exponentially suppressed by the subsequent stage
of inflation.

In the previous section we considered the simplest model where
$V(\phi) = 0$ for  $\phi > 0$. However, in general  $V(\phi)$ may
becomes  flat not at $\phi = 0$, but only  asymptotically, at
$\phi \gtrsim M_p$.  Such theories have become rather popular now
in relation to the theory of quintessence; for a partial list of
references see e.g. \cite{wetter}. In such a case the backreaction
of created particles may never turn the scalar field $\phi$ back
to $\phi = 0$. Therefore the decay of the particles $\chi$ may
occur very late, and one can have very efficient preheating for
any values of the coupling constants $g$ and $h$.

On the other hand, if the  $\chi$ particles are stable, and if the
field $\phi$ continues rolling for a very long time, one may
encounter a rather unusual regime. If the particle masses
$g|\phi|$  at some moment approach $M_p$, the $\chi$ particles may
convert to black holes and immediately evaporate.

Indeed, in conventional  quantum field theory, an elementary
particle of mass $M$ has a Compton wavelength $M^{-1}$ smaller
than its Schwarzschild radius $2 M/M_p^2$ if $M \gtrsim M_p$.
Therefore one may expect that as soon as   $m_\chi = g|\phi|$
becomes greater than $M_p$, each $\chi$ particle  becomes a
Planck-size black hole, which immediately evaporates and reheats
the universe.  If this regime is possible, it should   be avoided.
Indeed, black holes of Planck mass may produce similar amounts of
all kinds of particles, including gravitinos. Therefore  if
reheating occurs because of black hole evaporation, then we will
return to the gravitino problem again.

Thus, the best possibility is to consider those versions of the
instant preheating scenario which do not lead to the creation of
stable particles of Planckian mass. It may seem paradoxical that
one needs to be careful about this constraint. Several years ago
it would have seemed impossible to produce particles of mass
greater than $5 \times10^{12}$ GeV during the decay of an inflaton
field of mass $m_\phi \sim 10^{13} $ GeV. Here we consider a
nonperturbative mechanism  of preheating which may produce
particles 5 orders of magnitude heavier than $m_\phi$. It is
interesting that the mechanism of instant preheating discussed in
this paper works especially well in the context of NO models where
all other mechanisms are rather inefficient.

\bigskip
\section*{Acknowledgments}
It is a pleasure to thank R. Kallosh,  P.J.E. Peebles,  A. Van
Proeyen,  A. Riotto, A. Starobinsky, I. Tkachev, and A. Vilenkin
for useful discussions. This work was supported  by CIAR and by
NSF grant AST95-29-225. The work of G.F. and A.L. was also
supported   by NSF grant PHY-9870115. We are grateful to Nick
Pritzker and to the organizers of  the Pritzker Symposium in
Chicago where some of the results reported in this paper were
obtained.


\begin{references}
\bibitem{super} Note that if one takes $g > 10^{-3}$,   radiative
corrections to the effective potential may considerably change its
shape \cite{book}. However, this does not happen in supersymmetric
theories where the contributions of fermions and bosons nearly
cancel each other.
\end{references}
\end{document}